\documentstyle[12pt,aasms]{article}
\tighten

\begin{document}

\title{Angular Size in a Static Universe}

\author{David F Crawford}\
\affil{School of Physics A28,
University of Sydney, Sydney, N.S.W.  2006, Australia}

\begin{abstract}
In principle the geometry of the universe can be investigated by
measuring the angular size of known objects as a function of
distance.
Thus the distribution of angular sizes provides a critical test
of the stable and static model of the universe described by
Crawford (1991,1993) that has a simple and
explicit  relationship
between the angular size of an object and its redshift.
The result is that the agreement with observations of galactic
diameters and the size of double radio sources with the static
model is much better than the standard (Big Bang) theory without
evolution. However there is still a small discrepancy at large
redshifts that could be due to selection effects.
\end{abstract}

\keywords{cosmology: quasars: angular size: galaxies}

\section{Introduction}
The testing of cosmological theories by investigating the dependence
of angular size as a function of redshift has slowly gained impetus due to the
increasing number of objects with large redshifts.
Because size is such a simple concept it should be an excellent tool to
investigate the geometry of the universe.
However there is
a paradox in that the observational results appear to fit  a static
Euclidean model
where angular size is inversely proportional to redshift and they
cannot be reconciled with the standard
cosmological model without invoking evolution or some other
effect such as size dependence on magnitude together with
selection effects.  There is no general agreement
(\cite{NILSSON93}) on what the evolution or special effects
should be.
For a given set of data it is usually possible to find an evolution
function that provides a good fit.  However the freedom provided by the use of
an evolution function
means that the use of angular size data to test the standard model reduces to
arguments about whether the evolution function is reasonable or whether it
can be disentangled from other z dependent effects.
Nilsson {\it et al} argue that because of a power-size anticorrelation
their data is consistent with no cosmological evolution. The reality of
this power-size anticorrelation depends on the cosmological model and for
the standard model on what value of the deceleration parameter ($q_0$) is
chosen.

Here it will be shown that the angular size data is in agreement with the
stable static cosmology without requiring any evolutionary function.
My point is that not that the standard model with or without
evolution is wrong but that the data is also consistent with
a static cosmology that has no extra free parameters or functions.
The basic equations for the static cosmology are derived and
then the extensive double radio source
data given by Nilsson {\it et al.,} (1993) are analysed.
Except for selection effects the
agreement is excellent. Finally the static model is also compared with galaxy
size
data from Djorgovski and Spinrad  (1981) and radio data from
Kapahi (1987).

\section{The Cosmological Model}
The cosmological model described in Crawford (1991,1993)
has the
same geometry as the static Einstein universe (and incidentally the
same as a closed big bang model that is not expanding) which is
that for a three dimensional surface of a four dimensional
hyper-sphere.
In this theory the Hubble redshift is caused by a gravitational
interaction with the inter-galactic plasma (\cite{CRAWFORD87A}).
The redshift is a function of the distance and the
density of the plasma.   Provided the plasma density does not
vary significantly the redshift is a good measure of the distance
to an object.

Since the cosmological model is static and is not evolving it
obeys the {\em perfect cosmological principle} (\cite{BONDI48}) in
which there is both spatial and time isotropy.  This does not
prevent objects such as galaxies or quasars from evolving.  What it
does require is that their creation and evolution is independent
of their location in space and  in time.  Hence any sampling of
the universe over a sufficiently large scale should not detect
any variation in the average characteristics of the objects such
as size, luminosity or density.
The application of this principle to quasars, galaxies or radio sources
 requires that their
average linear size should be independent of their
redshift.
A critical test that would refute the static cosmology
is unequivocal evidence of a change in linear size with redshift
that is not due
to selection effects.

{}From Crawford (1991) the redshift ($z = \lambda_0/
\lambda_e - 1$) for a photon that has travelled a distance $r$ is
given by
\begin{equation}
 z = \exp\left(\frac{Hr}{c}\right) - 1,
\label{qso3}
\end{equation}
where $H$ is the Hubble constant.
For a static model $H$ is not the expansion velocity but a measure of the
observed redshift per unit distance.
One of the major results of the
model is to relate the Hubble constant to the universe's radius
$R$. From Crawford (1993) we get $R = \sqrt{2}c/\!H$ which
provides the basic relationship
\begin{equation}
r = \frac{R}{\sqrt{2}}\ln(1 + z).
\label{qso4}
\end{equation}

Let a source of radiation have a luminosity $L(\nu)$
($\mbox{W.Hz}^{-1}$) at the emission frequency $\nu$.  Then if
energy is conserved the observed flux density $S(\nu_0)$
($\mbox{W.m}^{-2}\mbox{.Hz}^{-1}$) at distance $r$  is the
luminosity divided by
the area which is
\[
 S(\nu_0) = \frac{L(\nu)}{4\pi R^2 \sin^2(r/\!R)}.
\]
However because of the gravitational interaction there is an
energy loss such that the received frequency $\nu_0$ is related
to the emitted frequency $\nu_e$ by equation~(\ref{qso3}) and is
\[
   \nu_0 = \nu_e \exp(\sqrt{2}r/\!R) = \nu_e/(1 + z).
\]
The loss in energy means that the observed flux density is
decreased by a factor of $1 + z$.  But there is an additional
bandwidth factor $d\nu_e = (1 + z)d\nu_0$ that tends to balance
the energy loss factor.  The balance is not perfect because the
source is observed at a different part of its spectrum from that
for a similar nearby source.  The correction for this spectral
offset is called the K-correction and
for radio sources the spectrum is often approximated by a
power law with
the form $L \propto \nu^{\alpha}$.
With this power law spectrum the
apparent luminosity including energy loss
and bandwidth corrections is
\begin{equation}
 S(\nu_0) = \frac{L(\nu_0)(1 + z)^\alpha}{4\pi R^2 \sin^2(r/
\!R)}.
\label{e2}
\end{equation}

For the geometry of the hyper-sphere the observed angular size of
an object (small angle approximation) with projected linear size $D$ at a
distance $r$ is given by
\begin{equation}
\theta = \frac{D}{R\sin{(r/\,R)}},
\label{e1}
\end{equation}
where $R$ is the radius of the universe.
Hence the angular size as a function of redshift is
\begin{equation}
\theta = \frac{D}{R\sin(\ln(1 + z)/\sqrt{2})}.
\label{e3}
\end{equation}
The angular size decreases with $z$ until $z = 8.22$ where it has
a broad minimum and then it increases with $z$ till it becomes
infinity at the antipole where $r = \pi R$ and $z = 84.02$.

\section{The observations}
The  measurement of angular size of an object over a large range
of distances has many problems. The first is that for an object to
act as a standard measuring rod it must correctly identified
at different distances. To a certain extent this can be avoided
by observing a large number of objects that are drawn from a
known statistical distribution. In the static model there is no
evolution so that it is required that this distribution is the
same (and hence the absolute size measurements are the same)
at all distances. The second problem is the difficulty of
measuring the size of objects such as galaxies with diffuse
edges. To a large extent this is overcome in the observations of
Djorgovski and Spinrad (1981)  who used
Petrosian (\cite{PETROSIAN76})
radii of bright elliptic galaxies.
They give the angular size and redshift for 25 galaxies.
Taking a different approach Kapahi (1987) measured the
separation of the lobes of radio galaxies. More recently this type of
measurement been repeated by
Nilsson {\it et al.,} (1993) who restricted their analysis to
Faranoff-Riley type II
double radio sources which have well defined edges so
that the angular size is easily determined.
Kellermann (1993) argued that compact radio sources observed
with VLBI should be free of many of the selection biases and has
provided results for a homogeneous group of bright compact radio
sources.  These compact objects have a nominal size of 41 pc
which is very small compared to a size of about 50 kpc for the
optical galaxies and 400 kpc for the double radio sources.
However his results do not agree with the other observations in
that they differ markedly from the usual approximately Euclidean
dependence on redshift.
A problem with these sources is that their size is not unambiguously
defined. Kellermann (1993) argues that they are not subject to systematic
evolutionary
effects and this explains the good agreement of their angular size  with the
standard model with $q_0 = 1/2$.
Since Kellermann's (1993) results differ greatly from the other data and
clearly
apply to  quite different structures they are not considered here.

\section{Analysis} The 540 double radio sources
(all Faranoff-Riley type II) listed
by Nilsson {\it et al.,} (1993) have been analysed using the static
cosmological model. They assembled a new large sample from the literature
using published radio maps to measure the angular size and
to determine the morphology or where
maps were not available they used the original data.
The values for the linear size in the static cosmology  were
computed by multiplying the linear sizes for the standard model
($H_0= \mbox{km.s}^{-1}\mbox{.Mpc}^{-1}$, $q_0 = 0$) given by Nilsson
{\it et al.,}
(1993,table(1)) by
the distance ratio defined by
\[
f_d = \frac{\sqrt{2}\sin(\ln(1+z)/\sqrt{2}))}{\frac{1}{2}\left[1 -
\frac{1}{(1+z)^2}\right]}.
\]
Similarly the values for the radio luminosity were computed using
\[
L_{static} = f_d^2 (1+z)^{-3} L_{standard}
\]
where the factor in $1+z$ removes the redshift corrections that
are not applicable in the static model.
Note that the {\em tired light} model discussed by Nilsson {\it et al.,}
 (1993)
and other authors differs from the static cosmology in that it
has a Euclidean geometry which gives a distance proportional to
$\ln(1+z)$ instead of proportional to  $\sqrt{2}\sin(\ln(1+z)/\sqrt{2})$.

\begin{figure}[htb]
\caption{Average Log(L/erg$\,\mbox{sec}^{-1}$) verses $z$
 for double radio sources
\label{magf}
for standard ($q_0 = 0$) and static cosmologies}
\epsfxsize=7cm

% GNUPLOT: LaTeX picture with Postscript
\setlength{\unitlength}{0.1bp}
\begin{picture}(3239,1943)(0,0)
\put(2264,536){\makebox(0,0)[r]{standard}}
\put(2264,636){\makebox(0,0)[r]{static}}
\put(1828,1892){\makebox(0,0){Figure 1}}
\put(1828,51){\makebox(0,0){z}}
\put(100,1021){%
%
\makebox(0,0)[b]{\shortstack{Log(L)}}%
%
}
\put(3056,151){\makebox(0,0){3}}
\put(2660,151){\makebox(0,0){2.5}}
\put(2264,151){\makebox(0,0){2}}
\put(1868,151){\makebox(0,0){1.5}}
\put(1471,151){\makebox(0,0){1}}
\put(1075,151){\makebox(0,0){0.5}}
\put(679,151){\makebox(0,0){0}}
\put(540,1792){\makebox(0,0)[r]{47}}
\put(540,1599){\makebox(0,0)[r]{46}}
\put(540,1407){\makebox(0,0)[r]{45}}
\put(540,1214){\makebox(0,0)[r]{44}}
\put(540,1022){\makebox(0,0)[r]{43}}
\put(540,829){\makebox(0,0)[r]{42}}
\put(540,636){\makebox(0,0)[r]{41}}
\put(540,444){\makebox(0,0)[r]{40}}
\put(540,251){\makebox(0,0)[r]{39}}
\end{picture}
\end{figure}
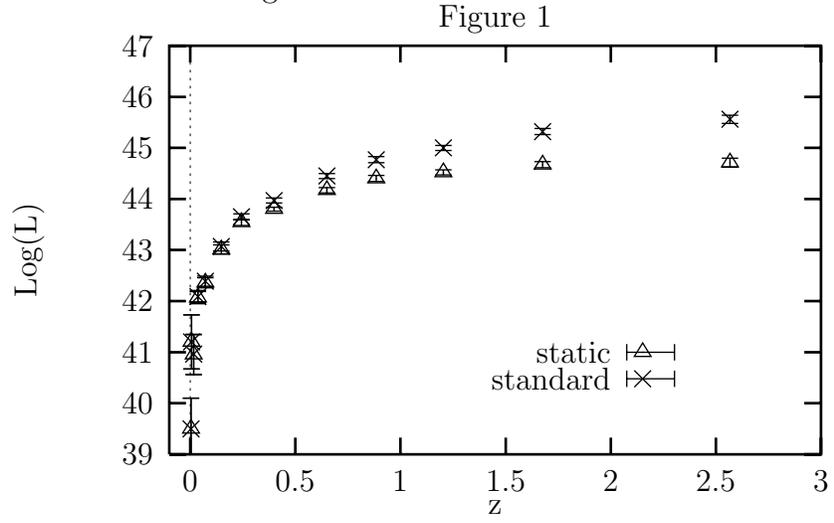

Nilsson {\it et al.,} (1993)  found that for the
standard cosmological model there is a strong dependence of
luminosity as a function of redshift.  Whereas the static model
shows a much smaller dependence of luminosity on redshift.
The average values of Log(L) for all objects are shown in
figure~(\ref{magf}) for both the standard model and the static
model.  Using the geometric mean of the luminosities (ie., the
average absolute magnitude) has the
desirable property that it is less sensitive to outliers than the
arithmetic mean.  The initial rise seen in figure~(\ref{magf}) for
both cosmological models is due to selection effects,
in particular it is due to the selection bias of the nearby
sources being weaker.  In this heterogeneous sample of radio
sources it is likely that most of them were discovered in flux
density limited surveys.  In addition there is usually a strong
correlation in radio surveys between the flux density limit and
the area of the survey in the sense that deep surveys with small
flux density limits are usually confined to a small area of the
sky whereas surveys that cover large areas invariably have larger
flux density limits.  The selection criteria is further
compounded by the different frequencies that were used in the
surveys.  The major effect of the flux density limits and the
area selection is the inclusion of an excessive number of nearby
weak sources which strongly bias the absolute magnitudes to
weaker values.  The computed values for the average absolute
magnitude and linear size (geometric average) as a function of
redshift are shown in table~(\ref{tmag}).  Because the data is
assembled from the literature without any control of selection
bias it is not clear whether the very slow increase in average
luminosity calculated on the static model for $z>0.5$ is a defect
in the model or is due to selection effects.
If the flux density selection limit is important it would be
evident as a deficiency of weak sources at large distances which
would show up as an increase in the average luminosity as $z$ increases.
The fact that the increase is so small for the static model suggests
that flux density selection limit is not very important and that
the static model is consistent with the data.

In order to see if there is a significant difference in the
luminosities of galaxies versus the luminosities of quasars
averages were taken for all objects with $z\geq 0.5$. For 69
galaxies the average of Log(L) was $44.60 \pm 0.05$ and for 219
quasars it was $44.40\pm 0.03$. Although the small difference is
statistically significant
it is more interesting that the absolute magnitudes of radio loud
quasars and galaxies are so similar.

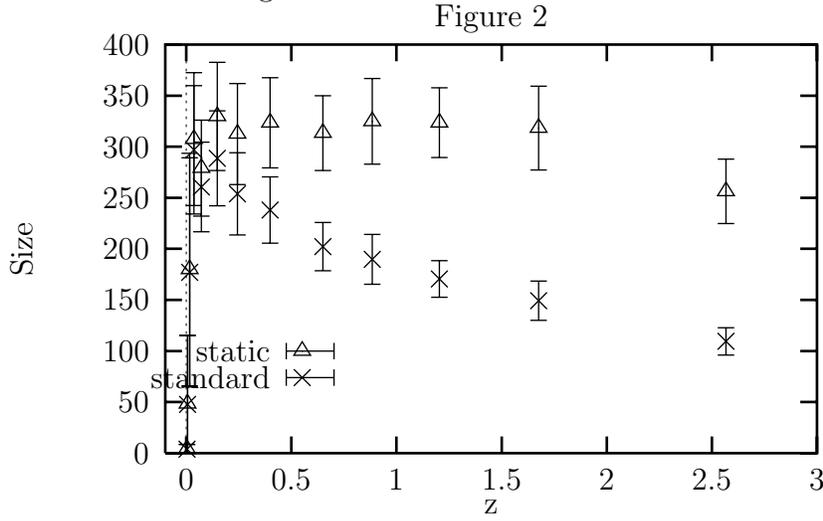
\begin{figure}[htb]
\caption{Average linear size (kpc) for double radio sources vs.
\label{suz}
redshift for standard ($q_0 = 0$) and static cosmologies}
\epsfxsize=7cm

% GNUPLOT: LaTeX picture with Postscript
\setlength{\unitlength}{0.1bp}
\begin{picture}(3239,1943)(0,0)
\put(996,536){\makebox(0,0)[r]{standard}}
\put(996,636){\makebox(0,0)[r]{static}}
\put(1828,1892){\makebox(0,0){Figure 2}}
\put(1828,51){\makebox(0,0){z}}
\put(100,1021){%
%
\makebox(0,0)[b]{\shortstack{Size}}%
%
}
\put(3056,151){\makebox(0,0){3}}
\put(2660,151){\makebox(0,0){2.5}}
\put(2264,151){\makebox(0,0){2}}
\put(1868,151){\makebox(0,0){1.5}}
\put(1471,151){\makebox(0,0){1}}
\put(1075,151){\makebox(0,0){0.5}}
\put(679,151){\makebox(0,0){0}}
\put(540,1792){\makebox(0,0)[r]{400}}
\put(540,1599){\makebox(0,0)[r]{350}}
\put(540,1407){\makebox(0,0)[r]{300}}
\put(540,1214){\makebox(0,0)[r]{250}}
\put(540,1022){\makebox(0,0)[r]{200}}
\put(540,829){\makebox(0,0)[r]{150}}
\put(540,636){\makebox(0,0)[r]{100}}
\put(540,444){\makebox(0,0)[r]{50}}
\put(540,251){\makebox(0,0)[r]{0}}
\end{picture}
\end{figure}

The (projected) geometric average linear size (for $H_0 = 50\,
\mbox{km.s}^{-1}\mbox{.Mpc}^{-1}$) of the radio sources is shown
in figure~(\ref{suz}) and in table~(\ref{tmag}) as a function of
redshift.
The small values at low redshifts reflect the same selection
effects that are apparent in the average luminosities and are of
little importance here.  Except for the last
point the static cosmology  projected linear sizes are essentially
constant for $z > 0.02$ which provides strong support for its validity.
{}From table~(\ref{tmag}) the average linear
size  for $z>2$ is $256\pm 32$ kpc to be compared with the
overall average of $315\pm 15$ kpc for $0.02 < z < 2.0$.
Although this is hardly statistically significant the
discrepancy gains credence from considering the galaxies and
quasars separately.
For $z>2.0$ there are 4 galaxies with an average linear size of
$251\pm 89$ kpc and 23 quasars with an average size of $257\pm
35$ kpc.  The agreement of these values suggests the the decrease
in linear size is real. However without knowing the selection criteria
for these sources it is too early to consider whether the decrease in size
for the highest redshift sources is  to  a defect in the
static cosmology or unknown selection effects.
This discrepancy in the highest redshift range is the strongest
evidence against the static model. But because it is the extreme
point and the numbers are small this discrepancy cannot be taken too seriously.

Kapahi (1987) concluded that based on angular size data
there was no difference between galaxies and quasars.
This is confirmed for the static model using the results
of Nilsson {\it et al.,} (1993) and is
consistent with them having similar absolute magnitudes. Using the
static cosmological model and taking a redshift range of $0.02
\leq z
\leq 2.0$ the  geometric average
linear size for 506 sources was $315\pm 15\,$ kpc.  In the same
range the average size for galaxies was $315\pm 22$ kpc and for
quasars it was $315 \pm 18$ kpc .
Closer examination of the results as a function of redshift show
that there is no significant difference between the linear size
of galaxies and that of quasars.

Since Nilsson {\it et al.,} (1993) found a  power-size anticorrelation
the correlation between the logarithm of the size and the
luminosity was determined. For all 540 sources the correlation coefficient was
$0.12$ which is significant at the 0.5\% level.  However if the sources with
sizes less than 30 kpc or greater than 2000 kpc are excluded the correlation
coefficient for 513 sources drops to  the insignificant value of 0.04
suggesting that the original correlation
was an artifact of some extreme values.
In any case the power-size anticorrelation is not found.
The conclusion is that  the occurrence of a significant power-size correlation
coefficient is clearly dependent on the cosmological model
that is used to determine the luminosities and linear sizes.

\section{Comparison of all data}
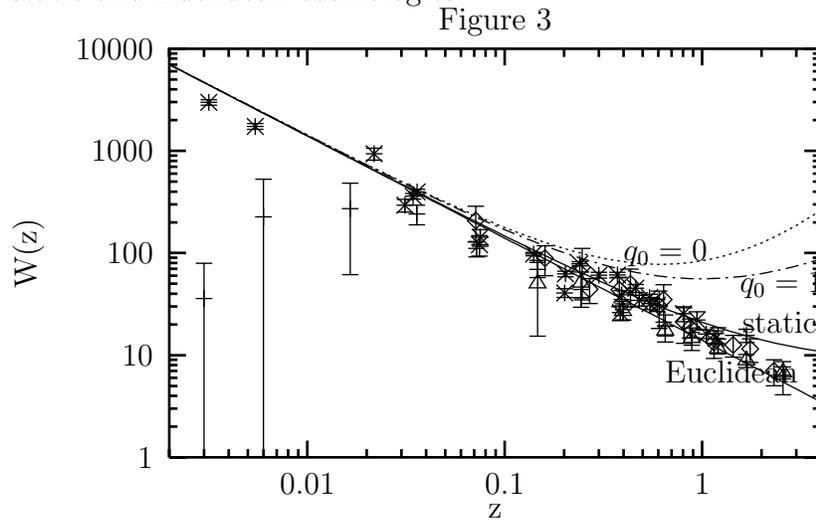
\begin{figure}[htb]
\caption{Angular size
verses redshift: $\diamondsuit$ - Kapahi (\protect 1987),
$\ast$ - Djorgovski and Spinrad (\protect 1981),
$\circ$ - Nilsson {\it et al.,} (\protect 1993) galaxies and
$\bullet$ - quasars. Curves are for
for standard ($q_0 = 0$, $q_0 = 1$), static and Euclidean cosmologies.
\label{kap}}
\epsfxsize=7cm

% GNUPLOT: LaTeX picture with Postscript
\setlength{\unitlength}{0.1bp}
\begin{picture}(3239,1943)(0,0)
\put(2469,574){\makebox(0,0)[l]{Euclidean}}
\put(2760,759){\makebox(0,0)[l]{static}}
\put(2750,900){\makebox(0,0)[l]{$q_0 = 1$}}
\put(2312,1020){\makebox(0,0)[l]{$q_0 = 0$}}
\put(1828,1892){\makebox(0,0){Figure 3}}
\put(1828,51){\makebox(0,0){z}}
\put(100,1021){%
%
\makebox(0,0)[b]{\shortstack{W(z)}}%
%
}
\put(2608,151){\makebox(0,0){1}}
\put(1864,151){\makebox(0,0){0.1}}
\put(1120,151){\makebox(0,0){0.01}}
\put(540,1792){\makebox(0,0)[r]{10000}}
\put(540,1407){\makebox(0,0)[r]{1000}}
\put(540,1022){\makebox(0,0)[r]{100}}
\put(540,636){\makebox(0,0)[r]{10}}
\put(540,251){\makebox(0,0)[r]{1}}
\end{picture}
\end{figure}

The angular size verses redshift data for all the radio and
optical objects from Djorgovski and Spinrad (1981),
Kapahi (1987)
and Nilsson {\it et al.,}  (1993)
are shown in figure~(\ref{kap}).  They have all
been normalized to have $\theta=200^{\prime\prime}$ at $z = 0.07$
so that at small values of redshift they should agree.  The four
curves are for Euclidean, static (equation~\ref{e3}) (static) and from
the standard theory with no evolution for $q_0 = 0$ and $q_0 =
1$.
Considering the selection effects and measurement difficulties
all the observations  are in reasonable agreement.
As already mentioned the observations differ markedly from the
standard model (without evolution) at large redshifts.  Nilsson
{\it et al.,}
(1993) provide a comprehensive discussion of the
various evolutionary models that have been suggested and other
alternatives such as a dependence of linear size on luminosity.
In fact Nilsson {\it et al} (1993) argue that their data can
be understood without cosmological evolution because of
a power-size anticorrelation.
Although some of the models have some physical plausibility there
is no general agreement on a suitable explanation that would
bring the standard model into agreement with the observations.
Whereas although the agreement with the static cosmological model
is very much better it is not perfect.  There appears to be a
small discrepancy (discussed earlier) with the Nilsson {\it et al.,}
(1993) results for the maximum redshift range
of $z>2$.  Note that some of the radio sources used by Kapahi
(1987) have had their redshifts determined from optical
magnitudes and in general his data have been superseded by that
from Nilsson {\it et al.,}
(1993).

\section{Conclusion}
It has been shown that the static cosmological model provides a
good fit to angular size data for large radio radio galaxies,
quasars and optical galaxies.
There is also support from the luminosity
data shown in figure~(\ref{magf}) for radio loud quasars and
galaxies, in that if the model is correct and selection effects
are small the luminosity should be independent of redshift.  What
is observed is apart from nearby sources where there are obvious
and expected selection effects the luminosity is almost constant.
There is a small anomaly in the angular size comparison at large
redshifts but since this could be due to section effects there is
no unequivocal evidence that would cause the static cosmology to
be rejected.

\section{Acknowledgements}
This work is supported by the Science Foundation for Physics
within the University of Sydney.

\clearpage
\begin{table}[htb]
\begin{center}
\caption{Log(L/erg$\,\mbox{sec}^{-1}$) and linear size as
 a function of redshift.
\label{tmag}}
\begin{tabular}{clrcc}
$z$ range& $\overline{z}$ &number & Log(L)  & Linear size (D)\\
\tableline
$0.002\leq z <0.005$ & 0.0030 &   2&  $39.50 \pm  0.60$ & $~~3\pm  ~~5$ \\
$0.005\leq z <0.01$  & 0.0060 &   3&  $41.20 \pm  0.53$ & $~48\pm  ~67$ \\
$0.01\leq z <0.02$   & 0.0165 &   7&  $40.95 \pm  0.39$ & $180\pm  114$ \\
$0.02\leq z <0.05$   & 0.036  &  23&  $42.07 \pm  0.12$ & $307\pm  ~65$ \\
$0.05\leq z <0.14$   & 0.072  &  54&  $42.36 \pm  0.10$ & $279\pm  ~47$ \\
$0.1\leq z <0.2$     & 0.147  &  45&  $43.01 \pm  0.09$ & $330\pm  ~53$ \\
$0.2\leq z <0.3$     & 0.244  &  44&  $43.54 \pm  0.06$ & $312\pm  ~49$ \\
$0.3\leq z <0.5$     & 0.398  &  74&  $43.80 \pm  0.05$ & $323\pm  ~44$ \\
$0.5\leq z <0.8$     & 0.65   &  80&  $44.17 \pm  0.05$ & $313\pm  ~37$ \\
$0.8\leq z <1.0$     & 0.88   &  48&  $44.40 \pm  0.06$ & $325\pm  ~42$ \\
$1.0\leq z <1.5$     & 1.20   &  79&  $44.52 \pm  0.05$ & $323\pm  ~34$ \\
$1.5\leq z <2.0$     & 1.68   &  54&  $44.67 \pm  0.06$ & $318\pm  ~41$ \\
$2.0\leq z <5.0$     & 2.57   &  27&  $44.71 \pm  0.08$ & $256\pm  ~32$ \\
\end{tabular}
\end{center}
\end{table}

\clearpage

\clearpage

\begin{thebibliography}{}

\bibitem[Bondi  \& Gold 1948]{BONDI48}
\reference Bondi, H.  \& Gold, T. 1948,
\mnras, 108, 252

\bibitem[Crawford 1987]{CRAWFORD87A}
\reference Crawford, D.~F. 1987,
Aust. J. Phys., 40, 449

\bibitem[Crawford 1991]{CRAWFORD91}
\reference Crawford, D.~F. 1991,
\apj, 377, 1

\bibitem[Crawford 1993]{CRAWFORD93}
\reference Crawford, D.~F. 1993,
\apj, 410, 488

\bibitem[Djorgovski  \& Spinrad 1981]{DJORGOVSKI81}
\reference Djorgovski, S.  \& Spinrad, H. 1981,
\apj, 251, 417

\bibitem[Kapahi 1987]{KAPAHI87}
\reference Kapahi, V.~K. 1987,
\reference in A. Hewitt, G. Burbidge,  \& L.~Z. Fang (eds.), Observational
  Cosmology, p. 251, IAU Symposium 124, Reidel, Dordrecht

\bibitem[Kellermann 1993]{KELLERMANN93}
\reference Kellermann, K.~I. 1993,
Nature, 361, 134

\bibitem[Nilsson et~al. 1993]{NILSSON93}
\reference Nilsson, K. Valtonen, M.~J. Kotilainen, J.,  \& Jaakkola, T. 1993,
\apj, 413, 453

\bibitem[Petrosian 1976]{PETROSIAN76}
\reference Petrosian, V. 1976,
Astrophys. Lett., 209, 1

\end{thebibliography}
\end{document}